%% file: document.tex
\pdfoutput=1
\newcommand*{\ATLASLATEXPATH}{}
\documentclass[cernpreprint,texlive=2016,txfonts,UKenglish]{\ATLASLATEXPATH atlasdoc}

\usepackage{atlaspackage}
\usepackage{atlasbiblatex}

\usepackage{atlasphysics}
\usepackage{siunitx}

\graphicspath{figures/}

\usepackage{document-defs}

\input{document-metadata}

\hypersetup{pdftitle={ATLAS document},pdfauthor={The ATLAS Collaboration}}

\begin{document}

\maketitle

\input{main_body}

\section*{Acknowledgements}
\input{sections/Acknowledgements}

\printbibliography

\newpage \input{atlas_authlist}


\end{document}

%% file: document-metadata.tex
\AtlasTitle{Measurement of the exclusive $\gamma \gamma \rightarrow \mu^+ \mu^-$ process in proton--proton collisions at $\sqrt{s}=13$\, \TeV{} with the ATLAS detector}

\author{The ATLAS Collaboration}

\AtlasRefCode{STDM-2016-13}

\PreprintIdNumber{CERN-EP-2017-151}

 \AtlasJournal{Phys. Lett. B}
 \AtlasJournalRef{Phys. Lett. B 777 (2018) 303}
 \AtlasDOI{10.1016/j.physletb.2017.12.043}

\AtlasAbstract{%
  The production of exclusive $\gamma \gamma \rightarrow \mu^+ \mu^-$ events in proton--proton collisions at a centre-of-mass energy of 13\, \TeV{} is measured with the ATLAS detector at the LHC, using data corresponding to an integrated luminosity of 3.2 fb$^{-1}$.
  The measurement is performed for a dimuon invariant mass of 12\, \GeV{} $<m_{\mu^+\mu^-}<$ 70\, \GeV{}.
The integrated cross-section is determined within a fiducial acceptance region of the ATLAS detector and differential cross-sections are measured as a function of the dimuon invariant mass.
The results are compared to theoretical predictions both with and without corrections for absorptive effects.

}

%% file: main_body.tex
\graphicspath{{./figures}}

\section{Introduction}
\label{sec:intro}
\input{sections/Introduction}

\section{ATLAS detector}
\label{sec:detector}
\input{sections/Detector}

\section{Data, simulated event samples and theoretical predictions}
\label{sec:modelling}
\input{sections/Theory}

\section{Event reconstruction, baseline selection and background estimation}
\label{sec:selection}
\input{sections/Event_selection}

\section{Exclusive selection}
\label{sec:extraction}

\input{sections/Exclusive_selection}

\section{Cross-section measurements}
\label{sec:cross_section}
\input{sections/Cross_section}

\section{Systematic uncertainties}
\label{sec:sys}
\input{sections/Systematics}

\section{Results}
\label{sec:results}
\input{sections/Results}

\section{Summary}
\label{sec:summary}
\input{sections/Summary}

%% file: sections/Introduction.tex
\graphicspath{{./figures/theory/}}

When proton--proton ($pp$) beams collide at the LHC, typically rare photon--photon induced ($\gamma\gamma$) interactions occur at perceptible rate and provide a unique opportunity to study high-energy electroweak processes~\cite{deFavereaudeJeneret:2009db}. Compared to other final states, the dilepton production is a standard candle process of the photon-induced production mechanism, thanks to its sizeable cross-section. Using $pp$ collisions at a centre-of-mass energy of $\sqrt{s}=7~\TeV{}$, measurements of $pp (\gamma\gamma) \rightarrow \mu^+\mu^- pp$ production (referred to as exclusive $\gmgmmumu$) were performed by the ATLAS and CMS collaborations~\cite{Chatrchyan:2011ci, Aad:2015bwa}.
The exclusive $\gmgmee$ process was also measured~\cite{Chatrchyan:2012tv,Aad:2015bwa}. 
A similar experimental signature has been used to study the $\gamma \gamma \rightarrow W^+ W^-$ reaction~\cite{Chatrchyan:2013foa, Khachatryan:2016mud, Aaboud:2016dkv}.

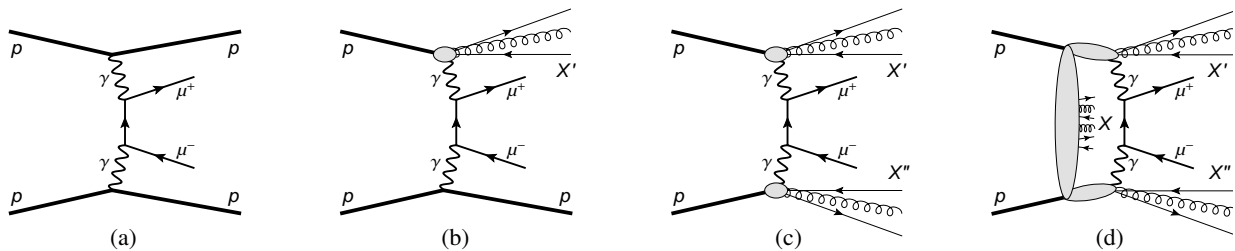
\begin{figure}[!htbp]
\begin{center}
 \includegraphics[width=1.0\textwidth]{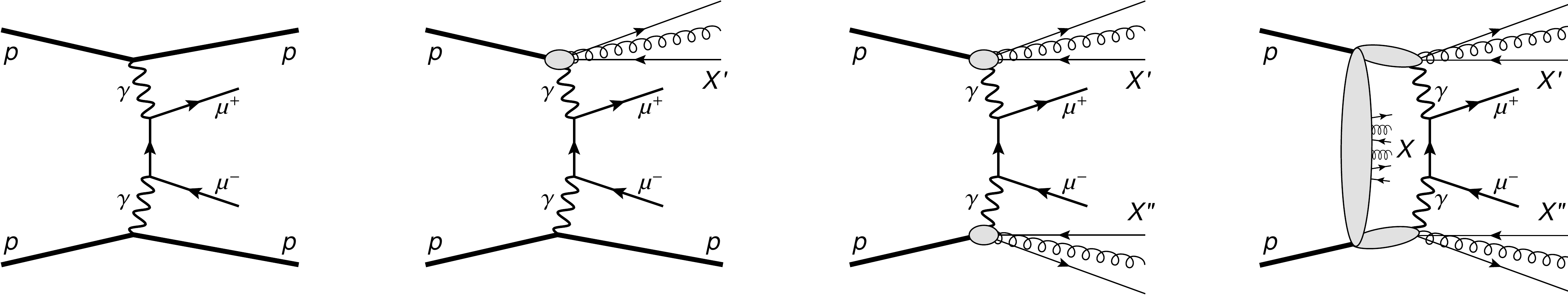}
 \put(-420,-3){{\footnotesize(a)}}
 \put(-296,-3){{\footnotesize(b)}}
\put(-171,-3){{\footnotesize(c)}}
\put(-51,-3){{\footnotesize(d)}}
\end{center}
 \caption{Schematic diagrams for (a) exclusive, (b) single-proton dissociative and (c) double-proton dissociative two-photon production of muon pairs in $pp$ collisions. The effect of additional interactions between the protons is shown in (d).}
\label{fig:diss_scheme}
\end{figure}

The exclusive $\gmgmmumu$ production process competes with the  two-photon interactions involving single- or double-proton dissociation due to the virtual photon exchange (Figure~\ref{fig:diss_scheme}~(a--c)). 
The electromagnetic (EM) break-up of the proton typically results in a production of particles at small angles to the beam direction, which can mimic the exclusive process.
However, the proton-dissociative processes have significantly different kinematic distributions compared to the exclusive reaction, allowing an effective separation of the different production mechanisms.


In general, the photon-induced production of lepton pairs contributes up to a few percent to the inclusive dilepton production at LHC energies~\cite{Aad:2014qja, Aad:2015auj, Aad:2016zzw}.

In order to reproduce the data, the calculations of such photon-induced reactions, in particular exclusive $\gmgmmumu$ production, need to take into account the proton absorptive effects~\cite{Aad:2015bwa}.
They are mainly related to additional gluon interactions between the protons (or proton remnants), shown in Figure~\ref{fig:diss_scheme}~(d), which take place  in addition to the QED process. 
The size of the absorption is not expected to be the same for exclusive and dissociative processes; it may also depend on the reaction kinematics. 
These effects lead to the suppression of exclusive cross-sections (typically around 10--20\%) by producing extra hadronic activity in the event besides the final-state muons.
Recent phenomenological studies suggest that the exclusive cross-sections are suppressed,
with a survival factor that decreases with mass~\cite{Dyndal:2014yea, Harland-Lang:2015cta}.

In  this  paper,  a measurement of  exclusive  dimuon  production  in  $pp$  collisions  at $\sqrt{s}=13$~\TeV{}~is presented for muon pairs with invariant mass $12~\GeV{} <m_{\mu^+\mu^-}< 70~\GeV{}$.
The  differential  cross-sections,  $\textrm{d}\sigma/\textrm{d}m_{\mu^+\mu^-}$,  are  determined within a fiducial acceptance region.
In the region $30~\GeV{}<m_{\mu^+\mu^-}<70~\GeV{}$, the  minimum transverse momentum of each muon
is required to be 10~\GeV{}.
For $12~\GeV{} <m_{\mu^+\mu^-}<30~\GeV{}$, the minimum muon transverse momentum is reduced to 6~\GeV{} by taking advantage of the lower trigger thresholds available by making additional requirements on muon-pair topology.
In addition, both muons are measured in the pseudorapidity range of $|\eta^{\mu}|<2.4$.
The measurements are compared to theoretical predictions both with and without corrections for absorptive effects.

%% file: sections/Detector.tex
The ATLAS experiment~\cite{Aad:2008zzm} at the LHC is a multi-purpose particle detector with a forward-backward symmetric cylindrical geometry and nearly 4$\pi$ coverage in solid angle.\footnote{ATLAS uses a right-handed coordinate system with its origin at the nominal interaction point in the centre of the detector and the $z$-axis coinciding with the axis of the beam pipe. The $x$-axis points from the interaction point to the centre of the LHC ring, and the $y$-axis points upward.  The pseudorapidity is defined in terms of the polar angle $\theta$ as $\eta=-\ln \tan(\theta/2)$, and $\phi$ is the azimuthal angle around the beam pipe with respect to the $x$-axis. 
The angular distance is defined as $\Delta R=\sqrt{(\Delta \eta)^2+(\Delta \phi)^2}$. 
The transverse momentum is defined relative to the beam axis.}
It consists of inner tracking devices surrounded by a superconducting solenoid, EM and hadronic calorimeters, and a muon spectrometer. 
The inner detector (ID) provides charged-particle tracking in the pseudorapidity region $|\eta|<2.5$ and vertex reconstruction. 
It comprises a silicon pixel detector, a silicon microstrip tracker, and a straw-tube transition radiation tracker.
The ID is surrounded by a solenoid that produces a 2\,T axial magnetic field. 
Lead/liquid-argon (LAr) sampling calorimeters provide EM energy measurements with high granularity. 
A hadron (steel/scintillator-tile) calorimeter covers the central pseudorapidity range $|\eta|<1.7$.
The end-cap and forward regions are instrumented with LAr calorimeters for both the EM and hadronic energy measurements up to $|\eta|=4.9$. 
The muon spectrometer (MS) is operated in a magnetic field provided by air-core superconducting toroids and includes tracking chambers
for precise muon momentum measurements up to $|\eta|=2.7$ and trigger chambers covering the range $|\eta|<2.4$.
 
A two-level trigger system~\cite{Aaboud:2016leb} selects the events used in the analysis. 
The first level is implemented in custom electronics, while the second trigger level is a flexible software-based system.

%% file: sections/Theory.tex
This analysis uses a data set of $pp$ collisions collected at a centre-of-mass energy $\sqrt{s}= 13~\TeV{}$ during 2015 under stable beam conditions. 
After applying data quality requirements, this data sample corresponds to an integrated luminosity of 3.2 fb$^{-1}$.

Calculations of the cross-section for exclusive $\gmgmmumu$ production in $pp$ collisions are based on the
Equivalent Photon Approximation (EPA)~\cite{Chen:1973mv,Budnev:1973tz}. 
The EPA relies on the property that the EM fields produced by the colliding protons
can be treated as a beam of quasi-real photons with a small virtuality of $Q^2 < 0.1~\GeV{}^2$.
This flux of equivalent photons is determined from the Fourier transform of the EM field of the proton, taking into account the EM form factors~\cite{Budnev:1974de}.
The cross-section for the reaction  $pp (\gamma\gamma) \rightarrow \mu^+\mu^- pp$ is calculated by convolving the respective photon fluxes with the elementary cross-section for the process $\gmgmmumu$.
The signal events for exclusive \gmgmmumu production were generated using the \textsc{\herwigSeven}~\cite{Bahr:2008pv, Bellm:2015jjp} Monte Carlo (MC) event generator, in which the  cross-section for the process is computed by combining the $pp$ EPA with the leading-order (LO) formula for \gmgmmumu.
It is found that the predictions for exclusive \gmgmmumu production from \textsc{Herwig} are identical to those from \textsc{\lpair}~4.0~\cite{Vermaseren:1982cz} generator.

The dominant background, photon-induced single-dissociative (S-diss) dimuon production (Figure~\ref{fig:diss_scheme}~(b)), was simulated using \textsc{\lpair}~4.0 with the  Brasse~\cite{Brasse:1976bf} and Suri--Yennie~\cite{Suri1972243} structure functions for proton dissociation. 
For photon virtualities $Q^2 < 5~\GeV{}^2$ and masses of the dissociating system $m_N<2~\GeV{}$, low-multiplicity states from the production and decays of $\Delta$ resonances are usually created. 
For higher $Q^2$ or $m_N$, the system decays into a variety of resonances, which produce a large number of forward particles.
The \textsc{\lpair}~package was interfaced to \textsc{JetSet}~7.408~\cite{Sjostrand:1993yb}, where the \textsc{Lund}~\cite{Andersson:1983ia} fragmentation model is implemented.

The \textsc{Herwig} and \textsc{Lpair} generators do not include any corrections to account for proton absorptive effects. Hence the normalisation of these MC samples is further constrained by a data-driven procedure, as described in Section~\ref{sec:cross_section}.

For double-dissociative (D-diss) reactions, \textsc{\pythia}~8.175~\cite{Sjostrand:2007gs} was used with the NNPDF2.3QED~\cite{Ball:2013hta} set of parton distribution functions (PDF).
The NNPDF2.3QED set uses LO QED and next-to-next-to-leading-order (NNLO) perturbative QCD (pQCD) calculations to construct the photon PDF, starting from the initial scale $Q^2_0 = 2~\GeV{}^2$. 
Additionally, two alternative PDF sets, CT14QED~\cite{Schmidt:2015zda} and LUXqed17~\cite{Manohar:2017eqh} are considered.
Depending on the multiplicity of the dissociating system, the default \textsc{\pythia}~8 string or mini-string fragmentation model was used for proton dissociation.
The absorptive effects in D-diss MC events are taken into account using the default multi-parton interactions model in~\textsc{\pythia}~8~\cite{Corke:2009tk}.

The NLO pQCD \textsc{Powheg-Box}~v2~\cite{Nason:2004rx, Frixione:2007vw, Alioli:2008gx, Alioli:2010xd} event generator was used with the CT10~\cite{Lai:2010vv} PDF to generate both the Drell--Yan (DY)~\Zstmumu~and~\Zsttautau~events. It was interfaced to \textsc{\pythia}~8.210~\cite{Sjostrand:2007gs} applying the AZNLO~\cite{Aad:2014xaa} set of generator-parameter values (tune) for the modelling of non-perturbative effects, including the CTEQ6L1~\cite{Pumplin:2002vw} PDF set.
The production of top-quark pair (\ttbar)~events was also modelled using
\textsc{Powheg-Box}, interfaced to \textsc{\pythia}~6.428~\cite{Sjostrand:2006za}.
The event generators used to model \Zstmumu, \Zsttautau and \ttbar reactions were interfaced to \textsc{\photos}~3.52~\cite{Golonka:2005pn, Davidson:2010ew} to simulate QED final-state radiation (FSR) corrections.

Multiple $pp$ interactions per bunch crossing (pile-up) were accounted for by overlaying simulated minimum-bias events, generated with \textsc{\pythia}~8.210 using the A2 tune~\cite{ATL-PHYS-PUB-2012-003},
 and reweighting the distribution of the average number of interactions per bunch crossing in MC simulation to that observed in data.
Furthermore, the simulated samples were weighted such that the $z$-position distribution of reconstructed $pp$ interaction vertices matches the distribution observed in data.
The ATLAS detector response was modelled using the GEANT4 toolkit~\cite{Agostinelli2003250, 2010EPJC} and the same event reconstruction as that used for data is performed.

The measured distribution of the exclusive \gmgmmumu process is compared with two models of absorptive corrections in Section~\ref{sec:results}.

In the finite-size parameterisation approach~\cite{Dyndal:2014yea}, the absorptive effects are embedded in the evaluation of the $\gamma\gamma$ luminosity, taking the photon energy and impact parameter dependence into account. 
A simple exponential form of the proton's transverse profile function, extracted from total and elastic $pp$ and $p\bar{p}$ cross-section data, is used to suppress the two-photon luminosity when the impact parameter of the $pp$ collision becomes small.
It determines the probability that no inelastic interaction producing additional hadrons in the final state occurs~\cite{Frankfurt:2006jp}.
 Moreover, only photons produced outside the proton with an assumed radius of $r_p=0.64$\,fm are allowed to initiate the two-photon process. This particular feature reflects the finite transverse size of the proton and leads to further suppression of the cross-section.

In the approach implemented in the \textsc{\superchic} event generator~\cite{Harland-Lang:2015cta}, the absorptive effects are included at the amplitude level differentially in the final-state momenta of scattered protons.
As a result, the suppression of the cross-section in general depends on the helicity structure of the $\gamma\gamma\rightarrow X$ sub-process and may also alter the kinematics of outgoing intact protons. 
Because some helicity amplitudes vanish for the $\gamma\gamma\rightarrow \mumu$ process in the limit of massless leptons, this effect plays a less significant role in the suppression of the $pp (\gamma\gamma) \rightarrow \mu^+\mu^- pp$ cross-section.
As in the model described above, the proton transverse profile function controls the reduction of the exclusive production cross-section when $pp$ collisions become central. It is fitted using a two-channel eikonal model to describe a range of total, elastic and diffractive $pp$ and $p\bar{p}$ data~\cite{Ryskin:2011qe}.

%% file: sections/Event_selection.tex
Events were selected online by a set of dimuon triggers with a muon \pT threshold of 6~\GeV{} or 10~\GeV{}, and dimuon invariant mass $10~\GeV{}<\Mmm<30~\GeV{}$ or $\Mmm>30~\GeV{}$ respectively.
 Triggers with the lower transverse momentum requirement were enabled for data-taking with an instantaneous luminosity below $1.2\times10^{34}$~cm$^{-2}$s$^{-1}$. 
 These triggers were designed to collect exclusive dimuon events by employing an additional selection on the transverse momentum of the dimuon system, $\pT^{\mu^+\mu^-}<2~\GeV{}$, to reduce contributions from DY and multijet production.

In each event, muon candidates are identified by matching complete tracks in the MS to tracks in the ID  and are required to be in the region $|\eta^{\mu}|<2.4$.
 The \textit{Medium} criterion, as defined in Ref.~\cite{Aad:2016jkr}, is applied to the combined tracks. 
The muons are required to be isolated using information from ID tracks and calorimeter energy
clusters in a cone around the muon using the so-called \textit{GradientLoose} criteria~\cite{Aad:2016jkr}.
For each muon, the significance of the transverse impact parameter,  defined by the transverse impact parameter ($d_0$)
 of a muon track with respect to the beam line divided by its estimated uncertainty ($\sigma_{d_0}$), is required to satisfy
$|d_0|/\sigma_{d_0}<3.0$.

Events are then required to have exactly one pair of oppositely-charged muons.
Muons are required to form a pair with an invariant mass of $12~\GeV{} <\Mmm< 30~\GeV{}$ or $\Mmm>30~\GeV{}$ with different $\pT^{\mu}$ conditions.
The offline $\pT^{\mu}$ requirements are identical to the trigger-level requirements, since the trigger efficiencies are found to be constant in the relevant $\pT^{\mu}$ range.
Each of the two muons must also be matched to one of the muons reconstructed by the trigger.

In order to select exclusive events, the average longitudinal impact parameter of the two leptons
is taken as the event vertex and is referred to as the dimuon vertex.
The longitudinal impact parameter of each muon track with respect to the dimuon vertex multiplied by the sine
of the track $\theta$ angle, is required to be less than 0.5 mm.

After these baseline selection requirements, $2.9\times 10^{6}$~dimuon candidates are found in the data. 

The background to the exclusive signal includes contributions from S-diss and D-diss \gmgmmumu~production, as well as $\Zstmumu$ or $\Zsttautau$, with less significant contamination due to  \ttbar~and multijet production.
S-diss and D-diss background contributions are estimated using MC simulation,
 with additional data-driven normalisation of the S-diss contribution as detailed in Section~\ref{sec:cross_section}.
The \Zg~and \ttbar~background  contributions are also estimated from simulation, 
and normalised using the respective inclusive cross-sections calculated at NNLO in perturbative QCD~\cite{Anastasiou:2003ds, Czakon:2011xx}.  
The background from $\gamma\gamma \rightarrow W^+W^-$ and $\gamma\gamma  \rightarrow \tau^+\tau^-$ processes contributes at a level below 0.2\% of the expected signal~\cite{Aaboud:2016dkv} and is therefore neglected.
The background contribution from $W$+jets production is also estimated to be negligible~\cite{Aad:2014qja}.
Scale factors are applied to the simulated samples to correct for the small differences between simulation and data in the  muon trigger, reconstruction and
identification efficiencies, as well as the momentum scale and resolution~\cite{Aad:2016jkr}. 
The efficiencies are measured using a tag-and-probe method combining results from $J/\psi\rightarrow\mumu$, $\Upsilon\rightarrow\mumu$, and $Z\rightarrow\mumu$ events to cover a large range in the muon transverse momentum.

The multijet background is determined using data-driven methods, similarly to the previous ATLAS exclusive dilepton measurement~\cite{Aad:2015bwa}.
It is extracted using same-charge muon pairs that satisfy the event selection criteria, except the requirement on muon charge.
The normalisation of the multijet background is determined by fitting the  invariant mass spectrum of the muon pair in the data to the sum of expected contributions, including MC predictions of the signal and the prompt muon backgrounds.

%% file: sections/Exclusive_selection.tex
A typical signature of exclusive  \gmgmmumu~events
is the absence of
charged-particle tracks, other than muon tracks~\cite{Aad:2015bwa, Aaboud:2016dkv}.
In contrast, inclusive background candidates (like DY or multijet) are produced with extra particles that originate from the emission and hadronisation of additional partons~\cite{Aad:2014jgf,Aad:2016ria}.
Therefore, in order to select exclusive \gmgmmumu~candidates,  a veto on additional charged-particle track activity is applied.
This vertex isolation requires no additional tracks with $\pt>400~\MeV{}$ and $|\eta|<2.5$ near the dimuon vertex with $|z_0^{\textrm{trk}}|<1$ mm, where $z_0^{\textrm{trk}}$ is the longitudinal impact parameter of track with respect to the dimuon vertex.
The value of 1 mm is optimised using the MC simulation and the expected signal significance. 
This value is identical to that used in Ref.~\cite{Aaboud:2016dkv}.

\begin{figure}[!htbp]
\begin{center}
 \includegraphics[width=.49\textwidth]{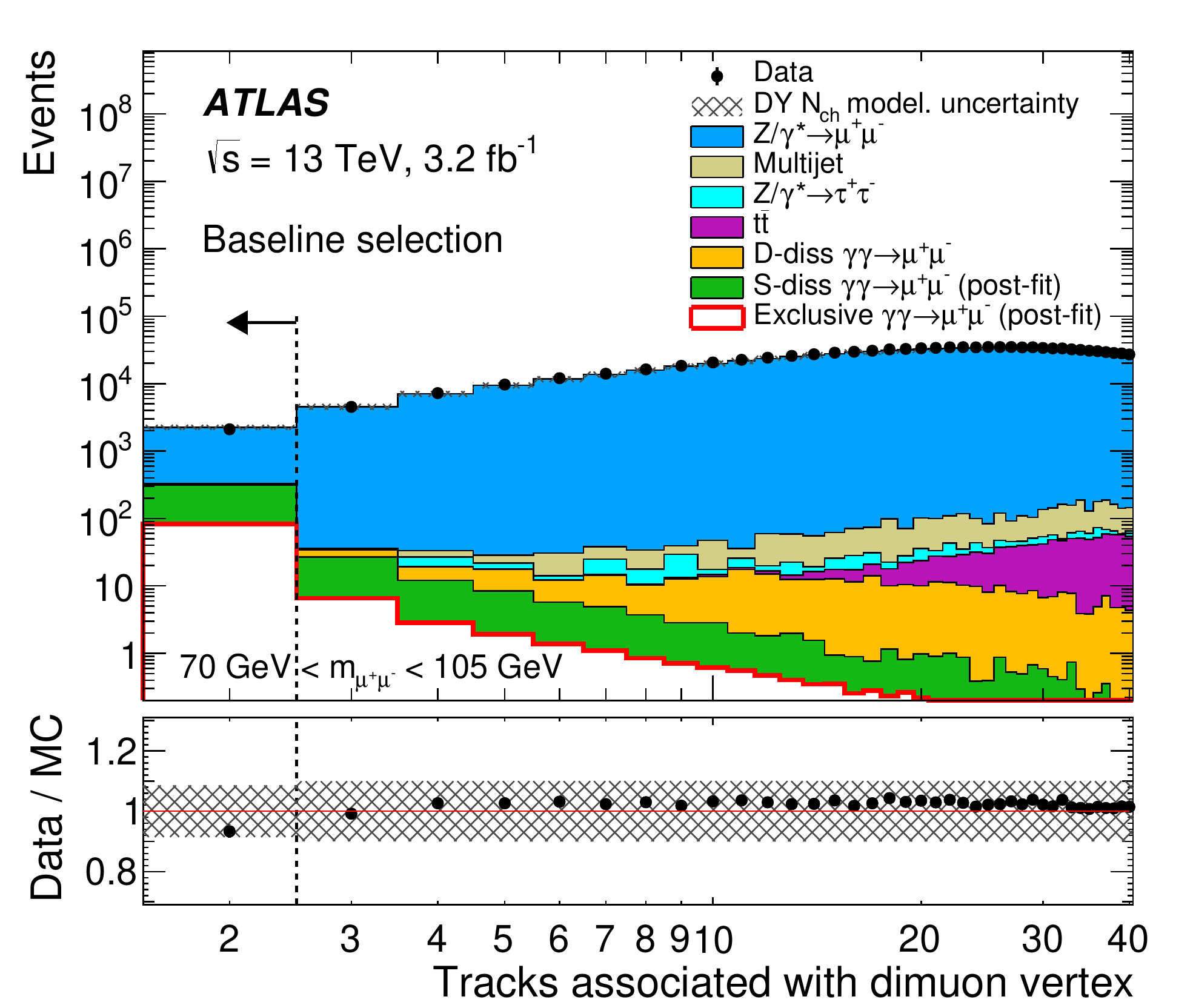}\put(-105,-8){{\footnotesize(a)}}
 \includegraphics[width=.49\textwidth]{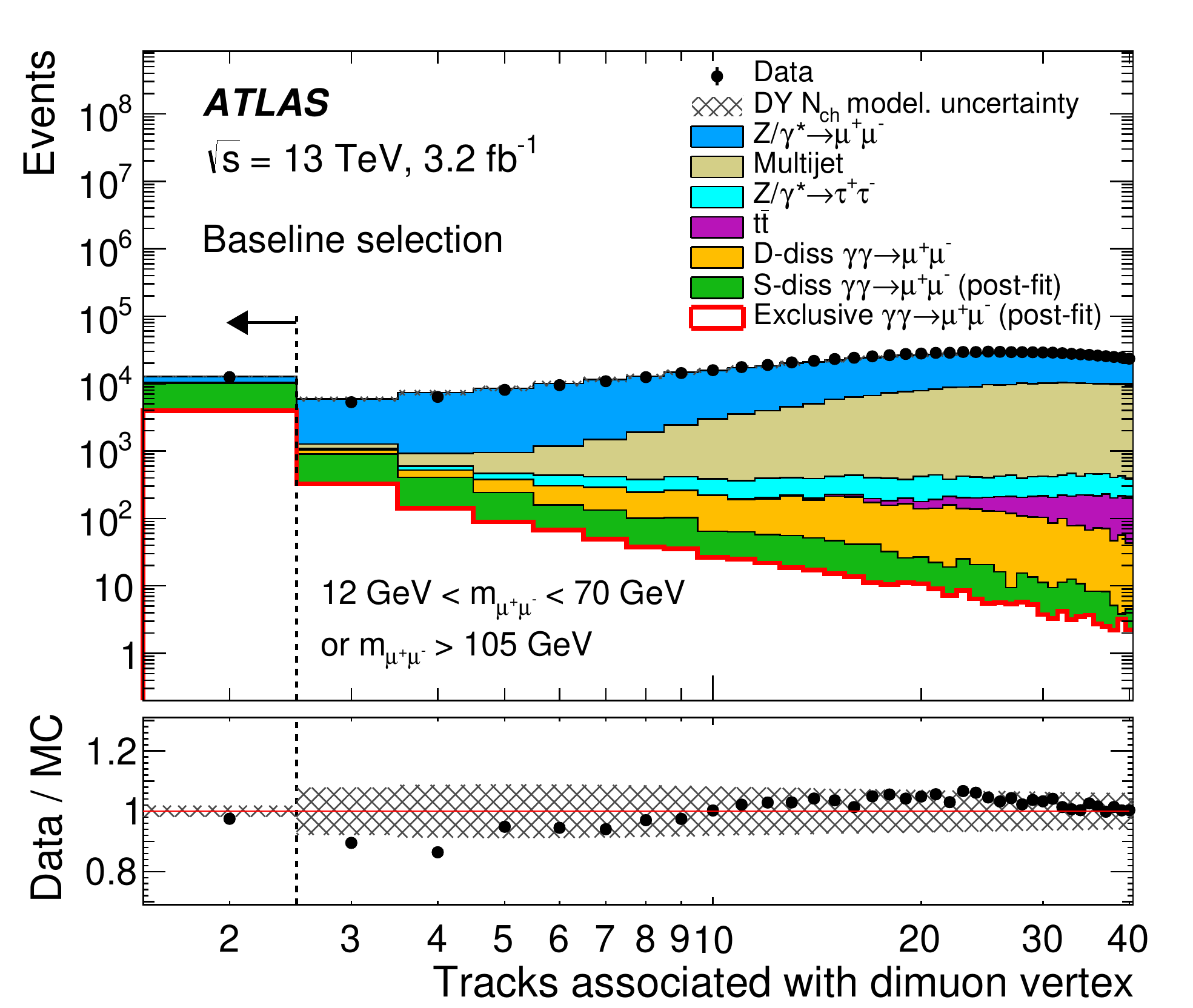}\put(-105,-8){{\footnotesize(b)}}
 
  \includegraphics[width=.49\textwidth]{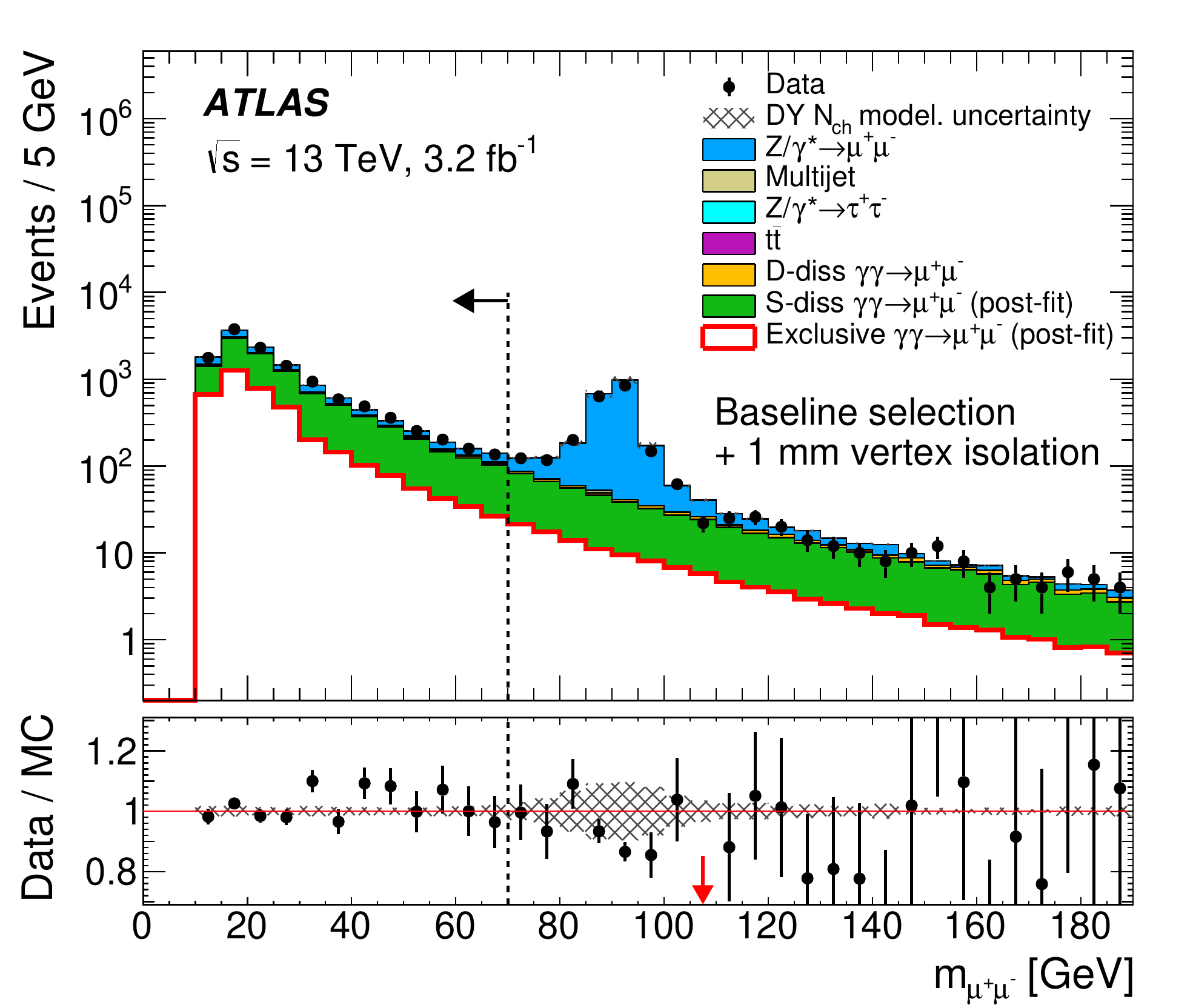}\put(-105,-8){{\footnotesize(c)}}
 \includegraphics[width=.49\textwidth]{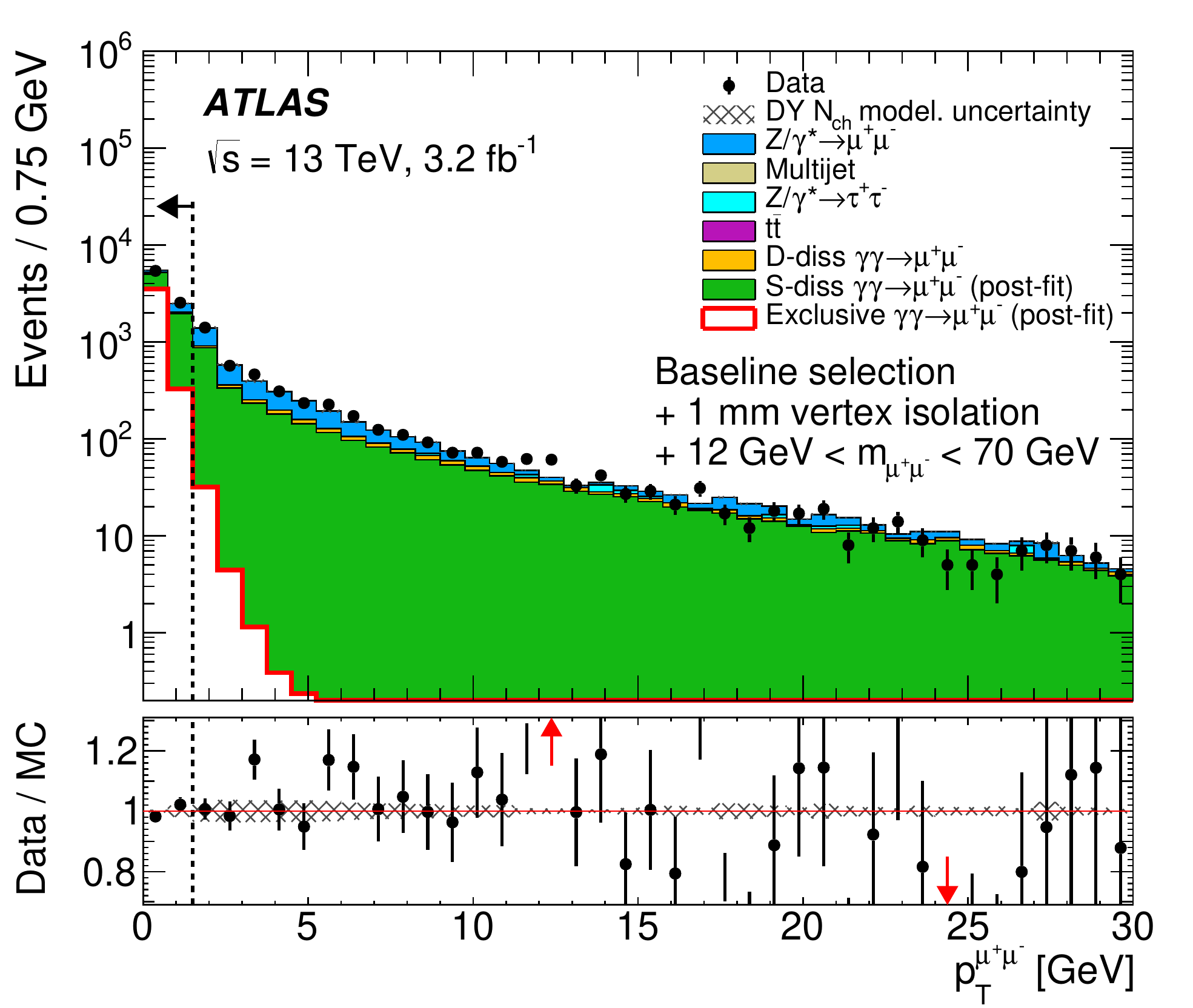}\put(-105,-8){{\footnotesize(d)}}
\end{center}
 \caption{ Illustration of event selection. 
 The distribution of the number of charged-particle tracks at detector level after applying the charged-particle reweighting procedure to DY MC simulation for (a) the $Z$-mass region and (b) the invariant mass range outside the $Z$-mass region.
(c) Dimuon invariant mass ($\Mmm$) distribution after applying 1 mm vertex isolation. (d) Transverse momentum of the dimuon system ($\pt^{\mu^+\mu^-}$) distribution after applying 1 mm vertex isolation and requiring $\Mmm<70~\GeV{}$.
 Data are shown as points with statistical error bars, while the histograms represent the expected signal and background levels. 
 The dashed vertical lines and arrows indicate the signal region selection.
 The uncertainty band indicates 10\% global uncertainty applied to DY simulation due to charged-particle multiplicity modelling. 
 The exclusive and S-diss yields are corrected using the fit procedure described in the text.
 The lower panels show the ratio of data to expected event yields.
 Red arrows in the lower panels indicate bins where the corresponding entry falls outside the plotted range.}
\label{fig:TrkCorrMCClos2}
\end{figure}

Following the procedure described in Refs.~\cite{Aad:2014jgf, Aad:2015bwa},
the shape of the charged-particle multiplicity distribution in simulated DY events is reweighted to match the spectrum observed in data.
The uncorrected $\Zst$ model overestimates the charged-particle spectrum observed in data by 50\% for low-multiplicity events.
In order to estimate the relevant weights, the events in the
$Z$-mass region (defined as $70~\GeV{} <m_{\mu^+\mu^-}<105~\GeV{}$) are used, since this region is expected to include a large DY contribution.
The distribution of the number of tracks associated with the dimuon vertex after applying the charged-particle reweighting procedure to DY simulation  is shown in Figure~\ref{fig:TrkCorrMCClos2} (a) for events in the $Z$-mass region.
A small mismodelling of this distribution is due to the contribution from fake tracks and secondary particles~\cite{Aad:2014jgf}, not taken into account in the correction procedure.
Similarly to Ref.~\cite{Chatrchyan:2012tb}, the underlying event activity in DY events is found to be independent of the dimuon invariant mass, down to $m_{\mu^+\mu^-} = 12~\GeV{}$.
For this reason, the same weights are applied to simulated DY events outside the $Z$-mass region (Figure~\ref{fig:TrkCorrMCClos2} (b)), and the description of charged-particle multiplicity is found to be satisfactory.
To cover differences observed between the data and simulation, a 10\% global uncertainty is assigned to DY MC simulation due to charged-particle multiplicity modelling.

\graphicspath{{./figures/}}

The invariant mass distribution of muon pairs for events satisfying the 1 mm vertex isolation is presented in Figure~\ref{fig:TrkCorrMCClos2} (c).
The contribution from DY events is further reduced by including only events with a dimuon invariant mass below 70~\GeV{}.
In order to further suppress the background from the S-diss process, the muon pair is required to have a transverse momentum, $\pt^{\mu^+\mu^-}$, below $1.5~\GeV{}$.
This is presented in Figure~\ref{fig:TrkCorrMCClos2} (d).

Table~\ref{tab:mumuCutflow} presents the effect of each step of the selection on the data and
the expected numbers of signal and background events.

\begin{table}[t!]
       \footnotesize
  \begin{center}
    \begin{tabular}{l|rrr|rrrrrr}
\hline \hline
&  &  & Total  &  &  & $\ensuremath{\Zzero/\gamma^{*}}$ & $\ensuremath{\Zzero/\gamma^{*}}$ &   &     \\
& Data & Signal & background & S-diss & D-diss & $\rightarrow\mumu$ & $\rightarrow\tautau$ & Multijet  & \ttbar    \\ 
\hline
     
Baseline selection & 2\,933\,384 & 5740 & 2\,897\,000 & 8640 & 8000 & 2\,268\,000 & 10\,900 & 590\,000 & 12\,200   \\

1 mm vertex isolation & 14\,759 & 4560 &  11\,100 & 6840 & 300 & 3900 & 30 & 50 & 0  \\

$\Mmm<70~\GeV{}$ & 12\,395 & 4420 &  8800 & 6420 & 300 & 2000 & 30 & 50 & 0  \\

$\ptmm<1.5~\GeV{}$ & 7952 & 4370 &  4300 & 3550 & 60 & 670 & 7 & 10 & 0  \\
\hline
    \end{tabular}
  \caption{ Effect of sequential selection requirements on the number of events observed in data, compared to the numbers of predicted signal and background events.
  }
  \label{tab:mumuCutflow}
  \end{center}
\end{table}

%% file: sections/Cross_section.tex
As in the previous ATLAS measurement~\cite{Aad:2015bwa}, the exclusive \gmgmmumu~contribution is extracted by performing a binned maximum-likelihood fit to the measured dimuon acoplanarity ($1-|\Delta\phi_{\mu^+\mu^-}|/\pi$) distribution.
The acoplanarity variable is not affected by the muon momentum scale and resolution uncertainties and provides a good separation of signal from background.
 Templates from MC simulation are used for the signal, DY, S-diss and D-diss processes. Contributions from other background sources are found to be negligible.
The fit determines two parameters: the expected number of signal events 
and the expected number of S-diss events.
The D-diss and DY contributions are fixed to their corresponding MC predictions in the fit procedure.

\begin{figure}[b!]
\begin{center}
 \includegraphics[width=.52\textwidth]{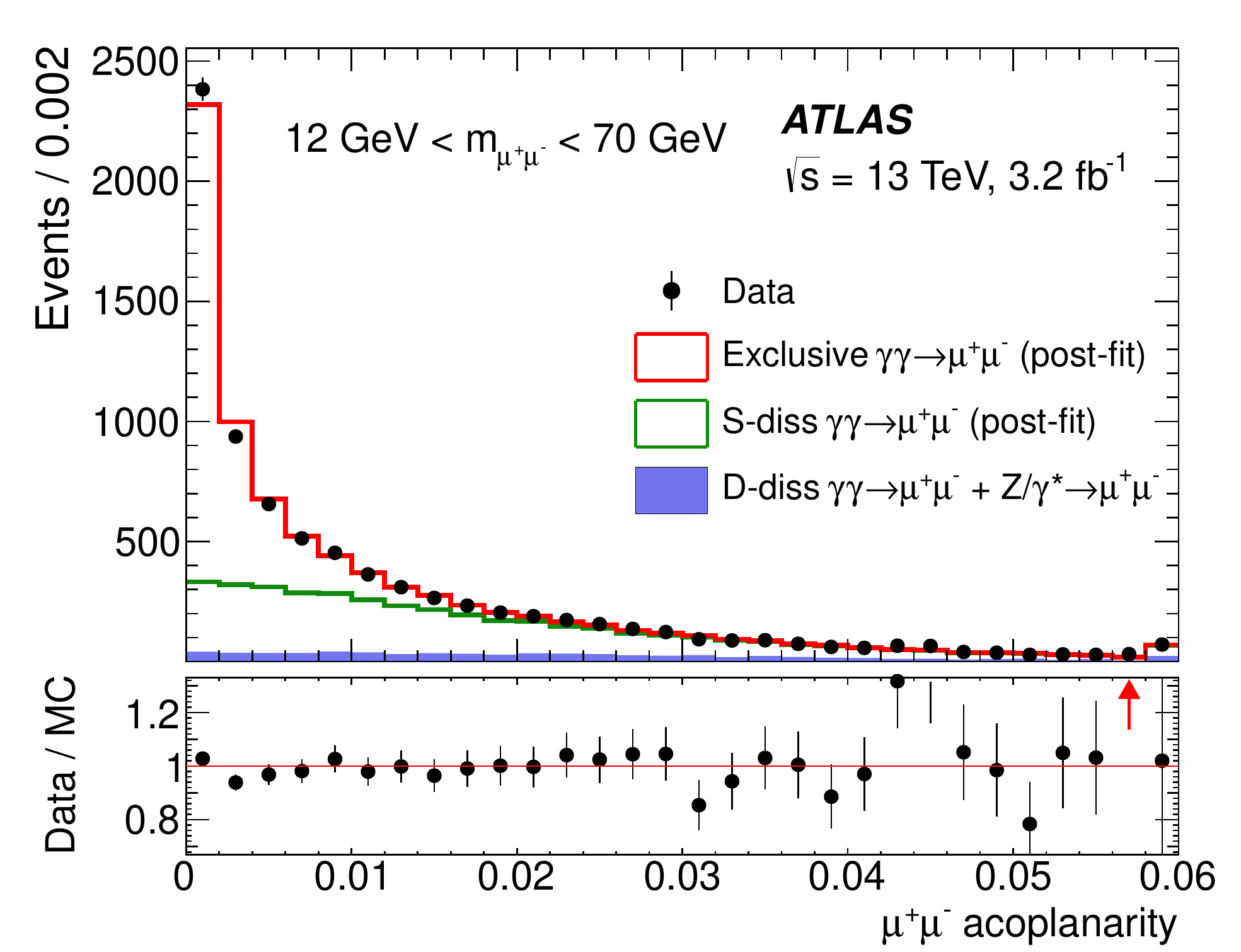}
\end{center}
 \caption{Dimuon acoplanarity distribution after signal selection requirements.
 Data are shown as points with
statistical error bars, while the histograms, in top-to-bottom order, represent the simulated exclusive signal, the S-diss and the sum of D-diss and DY backgrounds.
The exclusive and S-diss yields are determined from the fit described in the text.
 The last bin includes overflow events. The lower panel shows the ratio of data to the predicted distribution. 
 Red arrow in the lower panel indicates a bin where the corresponding entry falls outside the plotted range.}
\label{fig:fitMumuYield}
\end{figure}

The dimuon acoplanarity distribution in data overlaid with the result of the fit to the shapes from MC simulation is shown in Figure~\ref{fig:fitMumuYield} for the fiducial region.

The cross-section measurements  presented here correspond to the fiducial region defined in Table~\ref{tab:fidRegion}.
The fiducial cross-section for the process $pp (\gamma\gamma) \rightarrow \mu^+\mu^- pp$ is determined according to
\begin{linenomath*}
\begin{equation*}
\sigma^{\textrm{excl.~fid.}}_{\gmgmmumu} = \dfrac{N_{\textrm{excl.}}}{L_{\textrm{int}}\times C}~,
\end{equation*}
\end{linenomath*}
where $N_{\textrm{excl.}}$ is the total number of signal events extracted using the log-likelihood fit procedure, $L_{\textrm{int}}$ is the integrated luminosity of the data sample
and $C$ is the overall correction factor that accounts for efficiencies and resolution effects.
The $C$ factor is defined as the ratio of the number
of reconstructed MC signal events passing the selection to the number of generated MC
signal events satisfying the fiducial requirements.

\begin{table}[t!]
\footnotesize
  \begin{center}
    \begin{tabular}{lcc}
      \hline \hline 
	Invariant mass range  & $\pT^\mu$ requirement & $|\eta^\mu|$ requirement \\ \hline
    $12~\GeV{} <\Mmm< 30~\GeV{}$ & $>6~\GeV{}$ & $<2.4$ \\
    $30~\GeV{} <\Mmm< 70~\GeV{}$ & $>10~\GeV{}$ & $<2.4$ \\
      \hline
    \end{tabular}
  \end{center}
    \caption{Definition of the fiducial region for which the cross-sections are evaluated.}
    \label{tab:fidRegion}
\end{table}

The cross-section for exclusive dimuon production is also measured differentially in four bins of $m_{\mu^+\mu^-}$ from 12~\GeV{} to 70~\GeV{}. 
The bin widths are chosen to ensure purity above 90\%, where purity is defined as the fraction of reconstructed signal events in a given bin of $m_{\mu^+\mu^-}$ which were also generated in the same bin.
The differential measurement is unfolded for resolution effects using the signal simulation sample and a bin-by-bin correction procedure.
The differential fiducial cross-section as a function of the dimuon invariant mass is calculated as
\begin{linenomath*}
\begin{equation}
\left( \frac{\mathrm{d}\sigma^{\textrm{excl.}}_{\gmgmmumu}}{\mathrm{d}m_{\mumu}}\right)_i = \dfrac{N_{\textrm{excl.}}^i}{L_{\textrm{int}}\times  C_i \times (\Delta m)_i}~,
\label{eq:diffmass}
\end{equation}
\end{linenomath*}
where $N_{\textrm{excl.}}^i$ is the number of signal events recorded in the $i$-th invariant mass bin, $C_i$ is the correction factor in bin $i$ and $(\Delta m)_i$ is the width of the bin.

%% file: sections/Systematics.tex
The systematic uncertainties in the measurement enter the cross-sections determination through the calculation of the correction factors ($C_i$), the extracted number of signal events ($N_{\textrm{excl.}}^i$), or the estimation of $L_{\textrm{int}}$. 

The systematic uncertainties are classified as correlated or uncorrelated across the measurement bins.
The correlated contributions are propagated by the offset method in which the values from each source are coherently shifted upwards and downwards by one standard deviation and the magnitude of the change in the measurement is computed. 
The sign of the uncertainty corresponds to a one standard deviation upward shift of the uncertainty
source.
The uncorrelated sources are propagated using the pseudo-experiment  method  in  which  the  correction  factors  used  to  improve  the  modelling  of  data  by
the simulation are randomly shifted in an ensemble of pseudo-experiments according to the mean and
standard deviation of the correction factor.  
The resulting uncertainty in the measured cross-sections is determined from the variance of the measurements for the ensemble.  

\begin{description}
   \item[Muon-related sources:]
Uncertainties related to the muon trigger and selection efficiencies are studied using the $J/\psi\rightarrow\mumu$, $\Upsilon\rightarrow\mumu$ and $Z\rightarrow\mumu$ processes, and a tag-and-probe method~\cite{Aaboud:2016leb}.   
   
The muon trigger efficiency is estimated in simulation, with a dedicated data-driven analysis performed to obtain the simulation-to-data correction factors and the corresponding uncertainties.
The uncertainty in the correction factors $C_i$ in Eqn.~(\ref{eq:diffmass}) due to the statistical ($\delta^{\textrm{trig.}}_{\textrm{stat.}}$) and systematic ($\delta^{\textrm{trig.}}_{\textrm{syst.}}$) uncertainties in the trigger efficiency are around 0.3\% and 0.9\% respectively.

The muon selection efficiencies as determined from simulation are corrected with simulation-to-data correction factors, which have associated statistical  and systematic  uncertainties.
These contributions to the systematic uncertainty also affect $C_i$, and are denoted by $\delta^{\textrm{reco.}}_{\textrm{stat.}}$ and $\delta^{\textrm{reco.}}_{\textrm{syst.}}$ respectively.
The $\delta^{\textrm{reco.}}_{\textrm{stat.}}$ value is approximately 0.1\% and the $\delta^{\textrm{reco.}}_{\textrm{syst.}}$ value is around 1.0\%.

Uncertainties in the muon momentum calibration can cause a change of acceptance because of migration of events across the muon \pT thresholds and $m_{\mu^+\mu^-}$ boundaries.
They are obtained from a comparison of the $J/\psi$ and $Z$ boson invariant mass distributions in data and simulation~\cite{Aad:2016jkr}.
When propagated to the correction factors, this source is found to be below 0.5\%. 
This contribution is denoted by $\delta^{\textrm{sc./res.}}$.

\item[Vertex isolation efficiency:]

Since the dimuon vertex in each event occurs randomly within the Gaussian luminous region, the 1 mm dimuon vertex isolation efficiency is extracted from the data as follows: for each event $i$, a point $z_i$ is randomly chosen from a Gaussian distribution corresponding to the longitudinal shape of the luminous region, excluding a range of 20 mm centred about the dimuon vertex. 
This region is excluded to ensure any activity around point $z_i$ is unrelated to the dimuon vertex. 
In some events, the point $z_i$ is near tracks caused e.g. by pile-up interactions. The vertex isolation efficiency is defined as the fraction of events for which $z_i$ has no track within 1 mm. 
This efficiency, as measured in data, is compared with simulation in Figure~\ref{fig:MumuEffFromData} as a function of the number of reconstructed vertices ($N_{\textrm{vtx}}$). 
The average number of reconstructed vertices per event observed in data is approximately 9.
In general, good agreement between the data and simulation is observed, with a small systematic difference of 1--2\% observed in the region $8 < N_{\textrm{vtx}} < 12$, which impacts the $C_i$ by 1.1\% and this is taken as a systematic uncertainty.

It was also checked that the vertex isolation efficiency is well modelled by simulation for arbitrary choice of vertex isolation size.

Modelling of the muon impact parameter resolution may affect the vertex isolation efficiency and can give rise to additional systematic effects. 
This is estimated by varying the muon impact parameter resolution in simulation to match the shapes observed in data, and the impact on the cross-sections is found to be 0.3\%.

In total, the resulting uncertainty in the correction factors due to estimation of the vertex isolation efficiency is found to be $\delta^{\textrm{veto}}=1.2\%$.

\begin{figure}[t!]
\begin{center}
  \includegraphics[width=.47\textwidth]{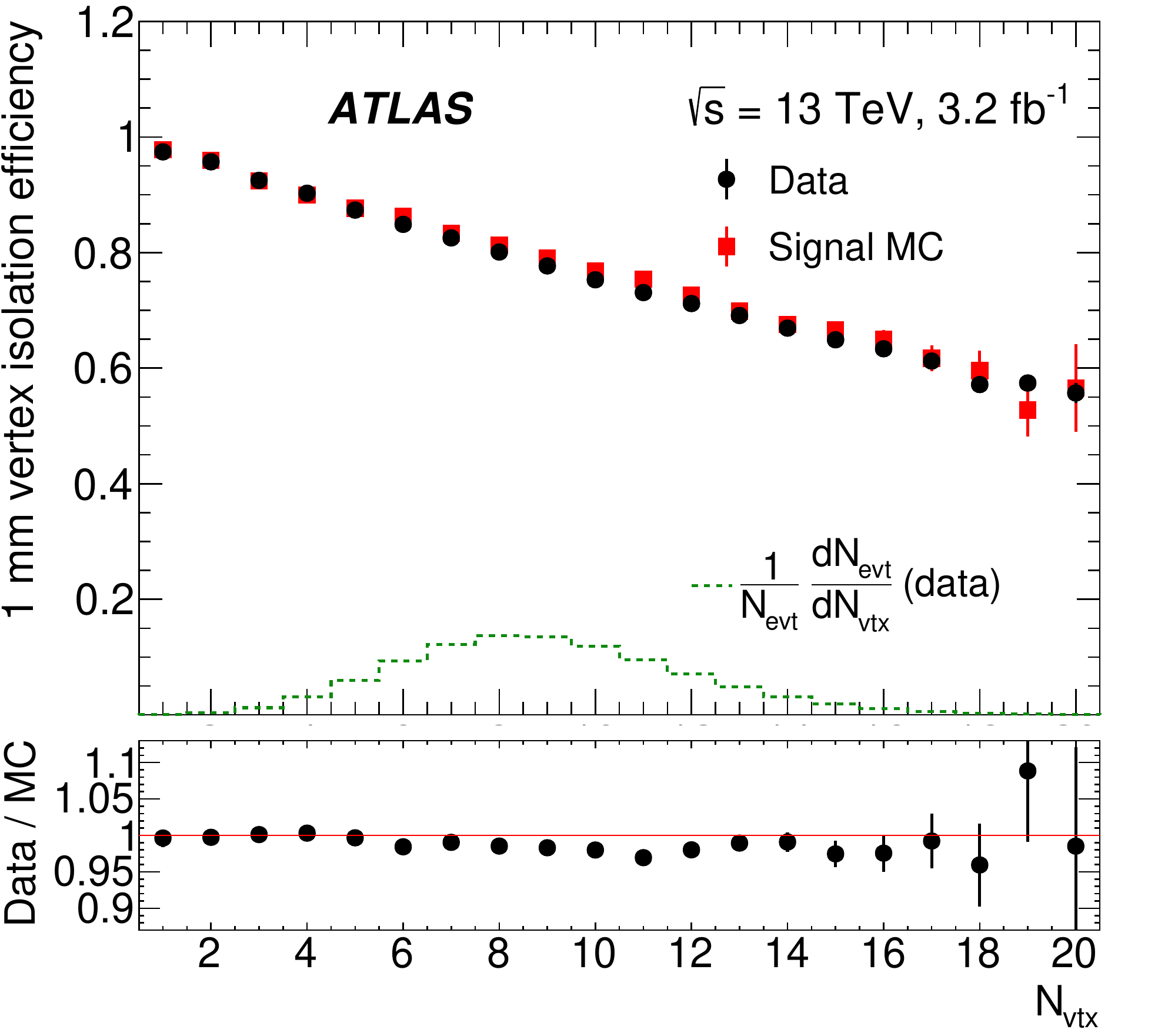}
\end{center}
 \caption{Dimuon vertex isolation efficiency for 1 mm requirement extracted from the data (black points) and signal MC simulation (red squares) as a function of the number of reconstructed vertices $N_{\textrm{vtx}}$. 
The distribution in the data is built according to the procedure described in the text.
The normalised $N_{\textrm{vtx}}$ distribution for data is shown as the dashed histogram.}
\label{fig:MumuEffFromData}
\end{figure}

\item[Pile-up description:]
The systematic effect related to the pile-up modelling is estimated  from the comparison between data and simulation of the \pT- and $\eta$-dependent density of tracks originating from pile-up, as in Refs.~\cite{Aad:2014jgf, Aad:2015bwa}.
The resulting uncertainty in $C_i$ is found to be $\delta^{\textrm{PU}}=0.5\%$ and  is fully correlated with the $\delta^{\textrm{veto}}$.

\item[Background:]
The uncertainty in the contribution of the DY process mainly accounts for disagreement between the data and simulation in  charged-particle multiplicity modelling (10\%). 
It also includes a 5\% contribution due to the PDF and scale uncertainties~\cite{Aad:2016naf}.
An overall normalisation uncertainty of 20\% is assigned to cover these effects.
Because of the similar shapes of the DY and S-diss \gmgmmumu~components in the fitted acoplanarity distribution, the uncertainty in the DY normalisation is partly absorbed by the S-diss contribution.
The 20\% uncertainty has typically a 0.7\% effect on the extracted number of signal events.

In order to estimate the D-diss \gmgmmumu~uncertainty, 
this contribution is varied according to the photon PDF uncertainties, defined at 68\% confidence level and evaluated using NNPDF2.3QED replicas~\cite{Ball:2013hta}.
The D-diss background uncertainty produces an uncertainty of 0.2\% in the cross-sections,
which is consistent with the full difference between the predictions obtained with the NNPDF2.3QED, CT14QED~\cite{Schmidt:2015zda} and LUXqed17~\cite{Manohar:2017eqh} central values.

The impact of these two background uncertainty sources is added in quadrature, yielding the uncertainty in $N_{\textrm{excl.}}^i$ ($\delta^{\textrm{bkg.}}$), which is less than 0.8\%.

\item[Template shape:]
The default signal acoplanarity template is constructed using bare EPA predictions from \textsc{Herwig}.
When using the acoplanarity templates from \textsc{SuperChic2} or from Ref.~\cite{Dyndal:2014yea}, the extracted number of signal events is lower by 2--3\%, which is taken as a systematic uncertainty.
The impact of the proton elastic form-factor modelling on the signal
acoplanarity template is evaluated in a similar way to Ref.~\cite{Aad:2015bwa} and takes into account differences between various parameterisations of proton EM form factors.
This has a 0.4\% effect on the extracted number of signal events.

The impact of the shape uncertainty in the S-diss
template is evaluated by varying the $\pT^{\mu^+\mu^-}$ requirement between \SI{1}{\GeV} and \SI{2}{\GeV}. 
The maximum deviation of $N_{\textrm{excl.}}^i$ from the nominal value
is observed to be 0.8\% and is taken as a systematic uncertainty.
In order to assign uncertainty due to the choice of proton structure functions in \textsc{\lpair}, an alternative set from Ref.~\cite{daSilveira:2014jla} is used. This impacts the $N_{\textrm{excl.}}^i$ by about 2.0\% and is taken as a systematic uncertainty.

When added in quadrature, these contributions are listed as $\delta^{\textrm{shapes}}$.

\item[LHC beam effects:]
The impact of the non-zero crossing angles of the LHC beams at the ATLAS interaction point is estimated by applying a Lorentz transformation to the generator-level lepton kinematics for signal MC events. 
This results in a negligible variation of the cross-sections.
The LHC beam energy uncertainty is estimated to be 0.1\%~\cite{Wenninger:2254678}.
It affects the cross-sections by less than 0.1\% and is considered to be a negligible effect.

\item[Unfolding method:]
The bin-by-bin correction used in the calculation of the cross-sections is compared to an iterative Bayesian unfolding technique~\cite{DAgostini:1994fjx}. The differences between these two approaches are found to be negligible.

\item[Luminosity:]
The uncertainty in $L_{\textrm{int}}$ is estimated to be $\delta^{\textrm{lumi.}}=2.1\%$. 
It is derived, following a methodology similar to that detailed in Ref.~\cite{Aaboud:2016hhf}, from a calibration of the luminosity scale using $x$--$y$ beam-separation scans performed in August 2015.

\item[Other cross-checks:]

To check the impact of MC modeling of neutral particles in the background processes, the analysis is repeated at generator level by requiring no extra neutral particle with $\pt>400~\MeV{}$ and $|\eta|<2.5$, in addition to the charged-particle exclusive selection. This extra requirement shows negligible impact on the analysis.

In similar generator-level studies, the $\pt$ threshold for charged particles is lowered to $100~\MeV{}$. The MC event yields for the dominant S-diss and a smaller D-diss background processes remain unchanged. For DY background the yield is suppressed by 80\%. No additional systematic uncertainty is, however, assigned as the DY contribution is constrained using $Z$-mass control region for a nominal selection with a total uncertainty of 20\%.

\end{description}

A summary of all systematic uncertainties is given in Table~\ref{tab:massDiffSyst}.

%% file: sections/Results.tex
\graphicspath{{./figures/results/}}

The fiducial cross-section is  measured to be
\begin{linenomath*}
$$\sigma_{\gmgmmumu}^{\textrm{excl.~fid.}} = 3.12~\pm~0.07~(\textrm{stat.}) ~\pm~0.14~(\textrm{syst.}) \ \pb~.$$
\end{linenomath*}
This value can be compared to the 
bare EPA predictions from \textsc{Herwig}, $\sigma_{\gmgmmumu}^{\textrm{EPA}} = 3.56~\pm~0.05$~pb,
to the EPA predictions corrected for absorptive effects using the finite-size parameterisation, $\sigma_{\gmgmmumu}^{\textrm{EPA,~corr.}} = 3.06~\pm~0.05$~pb, or to the \textsc{SuperChic2} predictions, $\sigma_{\gmgmmumu}^{\textrm{SC2}} = 3.45~\pm~0.05$~pb.
The theory uncertainties include uncertainties related to the knowledge of proton elastic form factors (1.5\%), and those originating from the higher-order electroweak corrections~\cite{Denner:1998tb} not included in the calculations (0.7\%).
These uncertainties are evaluated in a similar way as in Ref.~\cite{Aad:2015bwa}.

\begin{figure}[t!]
\begin{center}
 \includegraphics[width=.46\textwidth]{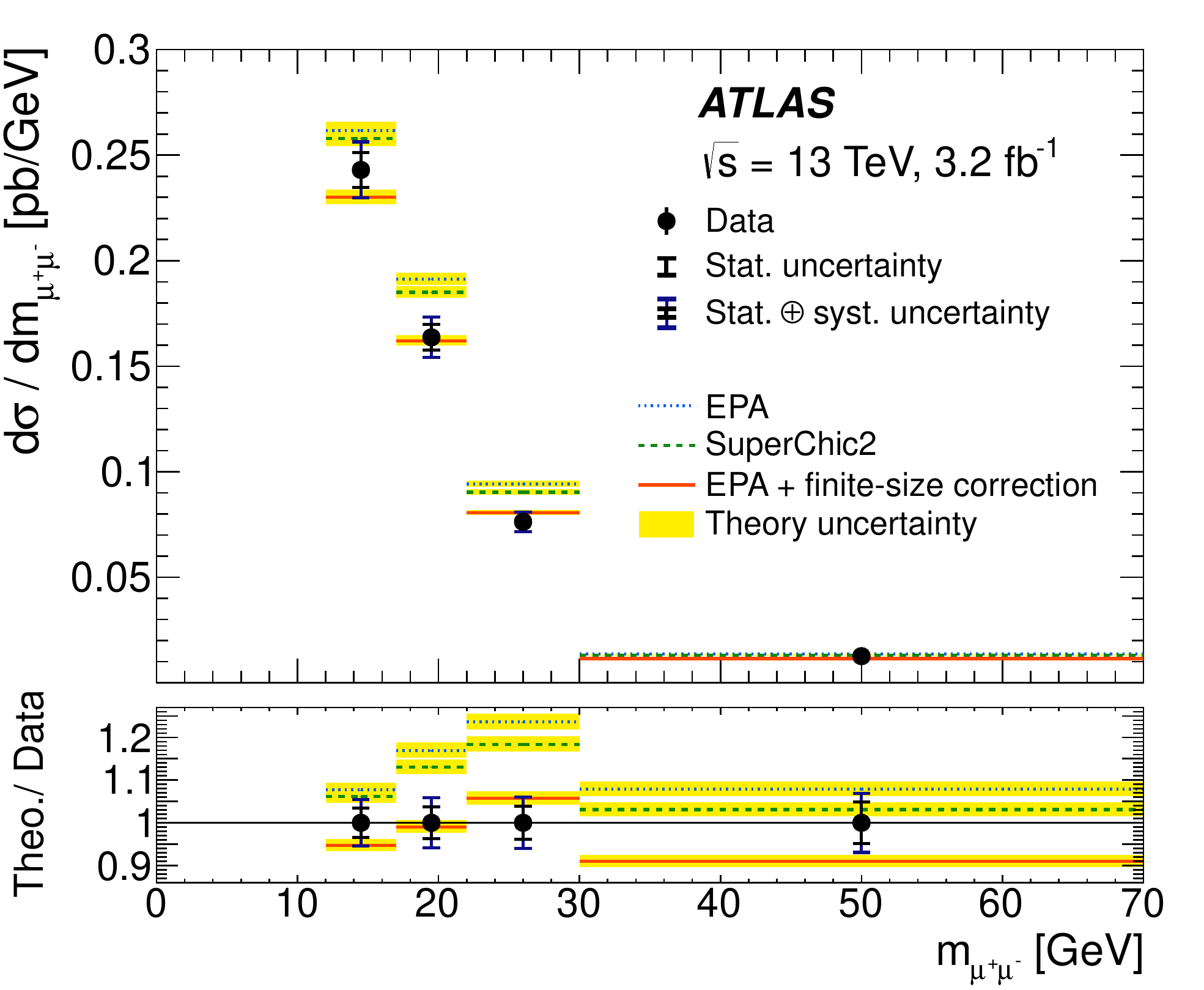}\put(-105,-8){{\footnotesize(a)}}
  \includegraphics[width=.477\textwidth]{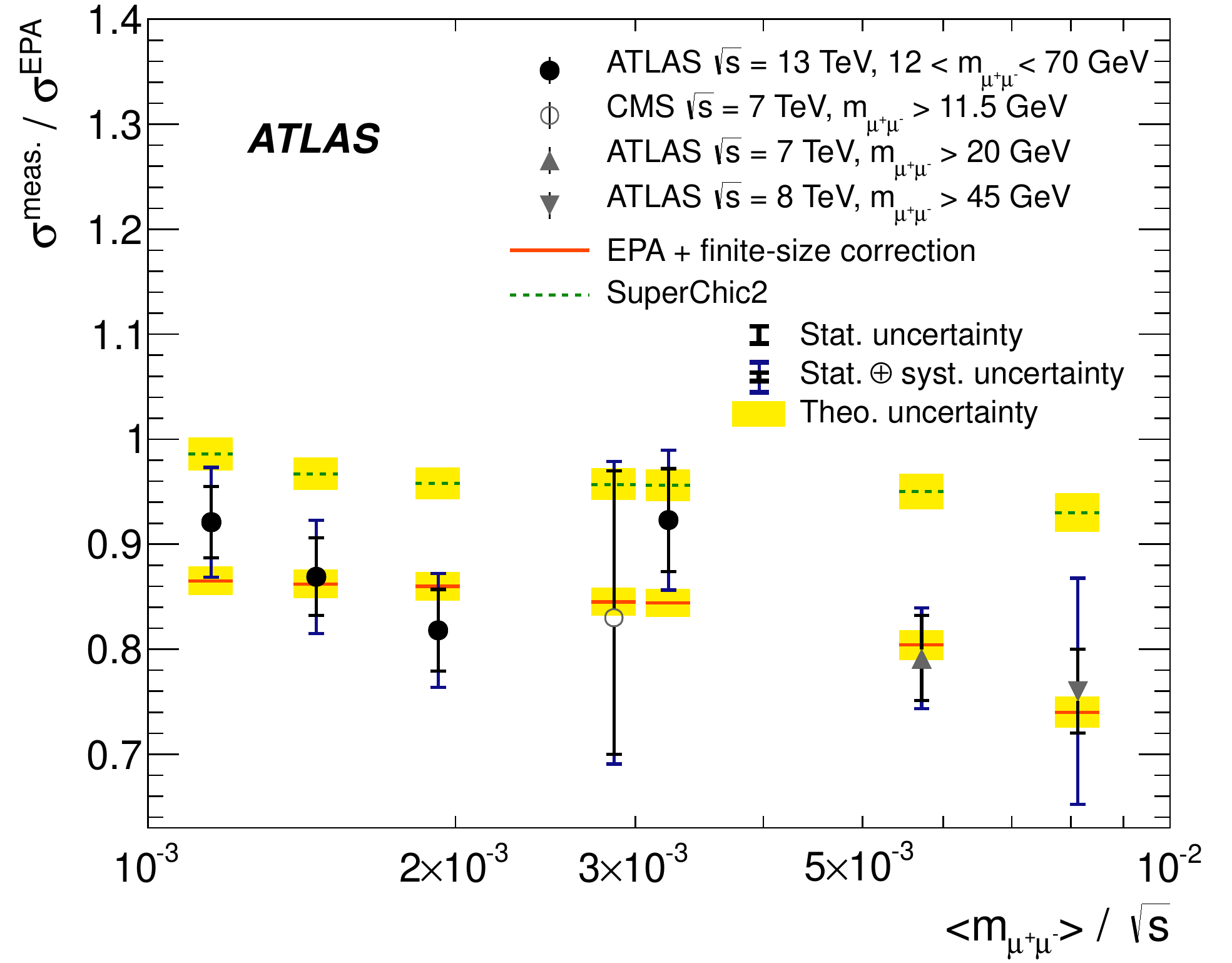}\put(-105,-8){{\footnotesize(b)}}
\end{center}
 \caption{(a) The exclusive \gmgmmumu differential fiducial cross-section measurements as a function of dimuon invariant mass $m_{\mumu}$. 
(b) Comparison of the ratios of measured and predicted cross-sections to the bare EPA calculations as a function of the average dimuon invariant mass scaled to the $pp$ centre-of-mass energy used. 
 Data (markers) are compared to various predictions (lines). Full circle markers represent the four mass points presented in this paper, while open circle, up-triangle and down-triangle depict the
 previous results obtained with $m_{\mumu}>11.5~\GeV{}$~\cite{Chatrchyan:2011ci}, $m_{\mumu}>20~\GeV{}$~\cite{Aad:2015bwa} and $m_{\mumu}>45~\GeV{}$~\cite{Aaboud:2016dkv} requirements on the dimuon invariant mass. The inner error bars represent the statistical uncertainties, and the outer bars represent the total uncertainty in each measurement. 
 The yellow bands represent the theoretical uncertainty in the predictions. 
 The bottom panel in (a) shows the ratio of the predictions to the data.}
\label{fig:MassDiff}
\end{figure}

The measured differential fiducial cross-sections as a function of dimuon invariant mass, together with the breakdown of the systematic uncertainties for the correlated and uncorrelated sources, are given in Table~\ref{tab:massDiffSyst}.
The comparison between the measured cross-sections and the theoretical predictions is shown in Figure~\ref{fig:MassDiff} (a).
The EPA predictions corrected for absorptive effects are in good agreement with the measured cross-sections.
The total systematic uncertainty of the measurement is dominated by shape modelling uncertainties, which can be reduced by tagging outgoing protons with dedicated forward detectors~\cite{Albrow:2014lrm, Adamczyk:2015cjy}.

It is expected that absorptive effects in two-photon interactions in $pp$ collisions depend on the proton energy fractions passed to the quasi-real photons (denoted by $x_1$ and $x_2$)~\cite{Dyndal:2014yea, Harland-Lang:2015cta}. 
Therefore, it is interesting to study the evolution of the survival factor, defined as the ratio of measured cross-section to the bare EPA predictions, as a function of the average dimuon invariant mass. 
Indeed, since $m_{\mumu}^2/s=x_1x_2$, where $s$ is the $pp$ centre-of-mass energy squared, the average values can be obtained:
\begin{linenomath*}
\begin{equation*}
\langle x\rangle ~\approx~\langle m_{\mumu}\rangle /\sqrt{s}~,
\end{equation*}
\end{linenomath*}
since at mid-rapidity ($y_{\mumu}\approx 0$) one has $x_1 \approx x_2$.

\begin{table}[t!]
\tiny
  \begin{center}
    \begin{tabular}{c|r@{$\pm$}cr@{$\pm$}c|ccc|cc|ccccccccc}
\hline \hline

& \multicolumn{2}{c}{} & \multicolumn{2}{c|}{} & & & & \multicolumn{2}{c|}{Uncorrelated} & \multicolumn{8}{c}{Correlated} \\

$m_{\mumu}$ & \multicolumn{2}{c}{$N_{\textrm{excl.}}^i$} & \multicolumn{2}{c|}{$C_i$} & $\textrm{d}\sigma/{\textrm{d}m_{\mumu}}$ & $\delta^{\textrm{stat.}}$ & $\delta^{\textrm{syst.}}$ & $\delta^{\textrm{trig.}}_{\textrm{stat.}}$ & $\delta^{\textrm{reco.}}_{\textrm{stat.}}$  & 
$\delta^{\textrm{trig.}}_{\textrm{syst.}}$ & $\delta^{\textrm{reco.}}_{\textrm{syst.}}$ & $\delta^{\textrm{sc./res.}}$ & $\delta^{\textrm{veto}}$ & $\delta^{\textrm{PU}}$ & $\delta^{\textrm{bkg.}}$ & $\delta^{\textrm{shapes}}$ &  $\delta^{\textrm{lumi.}}$ \\ 

[\GeV] &\multicolumn{3}{c}{} & & [pb/\GeV] & [\%] & [\%] & [\%] & [\%]& [\%]& [\%]& [\%]& [\%]& [\%]& [\%] & [\%]& [\%] \\
  
\hline     
12--17 & 1290 & 60 & 0.333  & 0.007 & 0.243 $\pm$ 0.013 & 3.4 & 4.3& 0.3 & 0.1   &  0.9 & 0.9 & -0.4 & -1.2 & -0.5 & 0.8 &  3.0  & 2.1 \\

17--22 & 1040 & 50 & 0.398  & 0.008 & 0.164 $\pm$ 0.010 & 3.7 & 4.5& 0.3 & 0.1   &  0.9 & 1.0 & -0.4 & -1.2 & -0.5 & 0.8 & 3.3  & 2.1 \\

22--30 & 830  & 40 & 0.428  & 0.009 & 0.076 $\pm$ 0.005 & 3.9 & 4.6& 0.2 & 0.1   &  0.9 & 1.0 & -0.2 & -1.2 & -0.5 & 0.6 & 3.5  & 2.1 \\

30--70 & 690  & 40 & 0.416  & 0.008 & 0.013 $\pm$ 0.001 & 4.9 & 4.9& 0.3 & 0.1   &  1.0 & 1.1 & -0.3 & -1.2 & -0.5 & 0.4 & 4.0 &  2.1 \\
\hline

12--70 & 3850 & 160 & 0.387 & 0.008 & 0.054 $\pm$ 0.003 & 2.1 & 4.5 & 0.3 & 0.1   &  0.9 & 1.0 & -0.3 & -1.2 & -0.5 & 0.8 & 3.3  & 2.1 \\
\hline
    \end{tabular}
  \caption{The measured exclusive \gmgmmumu differential fiducial cross-sections,  $\textrm{d}\sigma/\textrm{d}m_{\mu^+\mu^-}$. The extracted number of signal events ($N_{\textrm{excl.}}^i$) and correction factors ($C_i$) are also shown.
  The measurements are listed together with the statistical ($\delta^{\textrm{stat.}}$), and total systematic ($\delta^{\textrm{syst.}}$) uncertainties.
  In addition, the contributions from the individual correlated and uncorrelated systematic error sources are provided.
  The last row lists $\textrm{d}\sigma/\textrm{d}m_{\mu^+\mu^-}$ in the total fiducial region.
  The uncertainties in $N_{\textrm{excl.}}^i$ correspond to the combined statistical and systematic uncertainties. These are correlated across $m_{\mu^+\mu^-}$ bins.
  }
  \label{tab:massDiffSyst}
  \end{center}
\end{table}

\graphicspath{{./figures/}}

Figure~\ref{fig:MassDiff} (b) shows the evolution of the survival factor as a function of the average dimuon invariant mass scaled by a given $pp$ centre-of-mass energy. 
Exclusive two-photon production of muon pairs in $pp$ collisions at the LHC has been studied by the CMS experiment at $\sqrt{s} = 7~\TeV{}$ for $m_{\mumu}>11.5~\GeV{}$~\cite{Chatrchyan:2011ci}.
The ATLAS experiment measured exclusive production of muons at $\sqrt{s} = 7~\TeV{}$ in the region $\Mmm>20~\GeV{}$~\cite{Aad:2015bwa}.
Recently the production of exclusive \gmgmmumu at $\sqrt{s} = 8~\TeV{}$ was also studied by ATLAS in the context of exclusive $\gamma\gamma\rightarrow W^+W^-$ measurement~\cite{Aaboud:2016dkv}. 
The probed invariant mass region in this case is $\Mmm>45~\GeV{}$.
The $\langle m_{\mumu}\rangle$ for different measurements is calculated using the \textsc{Herwig} generator and corresponding fiducial region definitions.
The deviations from unity of the ratios of measured cross-sections to the bare EPA-based predictions from \textsc{Herwig} increase slightly with the energy scale $\langle m_{\mumu}\rangle /\sqrt{s}$.
This indicates that the size of the absorptive corrections tends to increase with $\langle m_{\mumu}\rangle /\sqrt{s}$.

The measurements are also compared to two model predictions that differ in the implementation of the absorptive corrections. 
While the finite-size parameterisation of absorptive effects describes the data reasonably well, mismodelling at the level of 10--20\% is observed with \textsc{SuperChic2}.
Moreover, at large masses, \textsc{SuperChic2} predicts less steeper decrease of the survival factor as a function of $\langle x\rangle$.
For example, the survival factor for fully exclusive $\gamma\gamma\rightarrow W^+W^-$ production at $\sqrt{s} = 13~\TeV{}$
is 0.82~\cite{Harland-Lang:2015cta} or 0.65~\cite{Dyndal:2014yea}, respectively.
A larger suppression of the EPA cross-sections in the finite-size approach is obtained by requiring that only photons outside the proton (with $r_p=0.64$\,fm) can initiate the exclusive photon-induced process.

%% file: sections/Summary.tex
A measurement of the cross-sections for exclusive \gmgmmumu production in $pp$ collisions at $\sqrt{s} = 13~\TeV{}$ with the ATLAS detector at the LHC is presented.
The measurement uses a data set corresponding to an integrated luminosity of $\SI{3.2}{\fb^{-1}}$.
The fiducial cross-section in the dimuon invariant mass range of $12~\GeV{} <m_{\mu^+\mu^-}<70~\GeV{}$
is measured to be $\sigma_{\gmgmmumu}^{\textrm{excl.~fid.}} = 3.12~\pm~0.07~(\textrm{stat.}) ~\pm~0.14~(\textrm{syst.}) \ \pb$.
 The differential cross-sections as a function of the dimuon invariant mass are also measured.

The cross-sections are compared to theoretical predictions which include corrections for absorptive effects.
The finite-size parameterisation of absorptive corrections provides a good description of the data, yielding $\sigma_{\gmgmmumu}^{\textrm{excl.~fid.}} = 3.06~\pm~0.05~\pb{}$.
It is observed that the absorptive corrections tend to increase with the energy fraction of protons passed to the initial-state photons. 
The precision of the measurement can be improved by using  dedicated forward proton detectors.

%% file: sections/Acknowledgements.tex

We thank CERN for the very successful operation of the LHC, as well as the
support staff from our institutions without whom ATLAS could not be
operated efficiently.

We acknowledge the support of ANPCyT, Argentina; YerPhI, Armenia; ARC, Australia; BMWFW and FWF, Austria; ANAS, Azerbaijan; SSTC, Belarus; CNPq and FAPESP, Brazil; NSERC, NRC and CFI, Canada; CERN; CONICYT, Chile; CAS, MOST and NSFC, China; COLCIENCIAS, Colombia; MSMT CR, MPO CR and VSC CR, Czech Republic; DNRF and DNSRC, Denmark; IN2P3-CNRS, CEA-DRF/IRFU, France; SRNSF, Georgia; BMBF, HGF, and MPG, Germany; GSRT, Greece; RGC, Hong Kong SAR, China; ISF, I-CORE and Benoziyo Center, Israel; INFN, Italy; MEXT and JSPS, Japan; CNRST, Morocco; NWO, Netherlands; RCN, Norway; MNiSW and NCN, Poland; FCT, Portugal; MNE/IFA, Romania; MES of Russia and NRC KI, Russian Federation; JINR; MESTD, Serbia; MSSR, Slovakia; ARRS and MIZ\v{S}, Slovenia; DST/NRF, South Africa; MINECO, Spain; SRC and Wallenberg Foundation, Sweden; SERI, SNSF and Cantons of Bern and Geneva, Switzerland; MOST, Taiwan; TAEK, Turkey; STFC, United Kingdom; DOE and NSF, United States of America. In addition, individual groups and members have received support from BCKDF, the Canada Council, CANARIE, CRC, Compute Canada, FQRNT, and the Ontario Innovation Trust, Canada; EPLANET, ERC, ERDF, FP7, Horizon 2020 and Marie Sk{\l}odowska-Curie Actions, European Union; Investissements d'Avenir Labex and Idex, ANR, R{\'e}gion Auvergne and Fondation Partager le Savoir, France; DFG and AvH Foundation, Germany; Herakleitos, Thales and Aristeia programmes co-financed by EU-ESF and the Greek NSRF; BSF, GIF and Minerva, Israel; BRF, Norway; CERCA Programme Generalitat de Catalunya, Generalitat Valenciana, Spain; the Royal Society and Leverhulme Trust, United Kingdom.

The crucial computing support from all WLCG partners is acknowledged gratefully, in particular from CERN, the ATLAS Tier-1 facilities at TRIUMF (Canada), NDGF (Denmark, Norway, Sweden), CC-IN2P3 (France), KIT/GridKA (Germany), INFN-CNAF (Italy), NL-T1 (Netherlands), PIC (Spain), ASGC (Taiwan), RAL (UK) and BNL (USA), the Tier-2 facilities worldwide and large non-WLCG resource providers. Major contributors of computing resources are listed in Ref.~\cite{ATL-GEN-PUB-2016-002}.

%% file: atlas_authlist.tex
\begin{flushleft}
{\Large The ATLAS Collaboration}

\bigskip

M.~Aaboud$^\textrm{\scriptsize 137d}$,
G.~Aad$^\textrm{\scriptsize 88}$,
B.~Abbott$^\textrm{\scriptsize 115}$,
O.~Abdinov$^\textrm{\scriptsize 12}$$^{,*}$,
B.~Abeloos$^\textrm{\scriptsize 119}$,
S.H.~Abidi$^\textrm{\scriptsize 161}$,
O.S.~AbouZeid$^\textrm{\scriptsize 139}$,
N.L.~Abraham$^\textrm{\scriptsize 151}$,
H.~Abramowicz$^\textrm{\scriptsize 155}$,
H.~Abreu$^\textrm{\scriptsize 154}$
% ...
\bigskip
\\
% ...
$^{12}$ Institute of Physics, Azerbaijan Academy of Sciences, Baku, Azerbaijan\\
% ...
$^{88}$ CPPM, Aix-Marseille Universit{\'e} and CNRS/IN2P3, Marseille, France\\
% ...
$^{115}$ Homer L. Dodge Department of Physics and Astronomy, University of Oklahoma, Norman OK, United States of America\\
% ...
$^{119}$ LAL, Univ. Paris-Sud, CNRS/IN2P3, Universit{\'e} Paris-Saclay, Orsay, France\\
% ...
$^{137}$ $^{(a)}$ Facult{\'e} des Sciences Ain Chock, R{\'e}seau Universitaire de Physique des Hautes Energies - Universit{\'e} Hassan II, Casablanca; $^{(b)}$ Centre National de l'Energie des Sciences Techniques Nucleaires, Rabat; $^{(c)}$ Facult{\'e} des Sciences Semlalia, Universit{\'e} Cadi Ayyad, LPHEA-Marrakech; $^{(d)}$ Facult{\'e} des Sciences, Universit{\'e} Mohamed Premier and LPTPM, Oujda; $^{(e)}$ Facult{\'e} des sciences, Universit{\'e} Mohammed V, Rabat, Morocco\\
% ...
$^{139}$ Santa Cruz Institute for Particle Physics, University of California Santa Cruz, Santa Cruz CA, United States of America\\
% ...
$^{151}$ Department of Physics and Astronomy, University of Sussex, Brighton, United Kingdom\\
% ...
$^{154}$ Department of Physics, Technion: Israel Institute of Technology, Haifa, Israel\\
$^{155}$ Raymond and Beverly Sackler School of Physics and Astronomy, Tel Aviv University, Tel Aviv, Israel\\
% ...
$^{161}$ Department of Physics, University of Toronto, Toronto ON, Canada\\
% ...
$^{a}$ Also at Department of Physics, King's College London, London, United Kingdom\\
$^{b}$ Also at Institute of Physics, Azerbaijan Academy of Sciences, Baku, Azerbaijan\\
$^{c}$ Also at Novosibirsk State University, Novosibirsk, Russia\\
$^{d}$ Also at TRIUMF, Vancouver BC, Canada\\
$^{e}$ Also at Department of Physics {\&} Astronomy, University of Louisville, Louisville, KY, United States of America\\
% ...
$^{*}$ Deceased
\end{flushleft}
